\documentclass[prd,showpacs,nofootinbib,tightenlines,preprintnumbers,%
superscriptaddress,twocolumn]{revtex4}
\usepackage{epsfig,graphics}
\usepackage{amssymb,amsmath,amsthm,graphicx,psfrag}

\newcommand{\bea}{\begin{eqnarray}}
\newcommand{\eea}{\end{eqnarray}}
\newcommand{\beq}{\begin{equation}}
\newcommand{\eeq}{\end{equation}}
\newcommand{\Fig}[1]{Fig.~\ref{#1}}

\newcommand{\Sec}[1]{Sec.~\ref{#1}}
\newcommand{\Eq}[1]{Eq.~(\ref{#1})}

\begin{document}

\preprint{HU-EP-06/03}
\preprint{ITEP-LAT/2006-01}

\title{Calorons and monopoles from smeared $SU(2)$ lattice fields at
non-zero temperature}

\author{E.-M. Ilgenfritz}
\affiliation{Institut f\"ur Physik, Humboldt-Universit\"at zu Berlin,
Newtonstr. 15, D-12489 Berlin, Germany}
\author{B. V. Martemyanov}
\affiliation{Institute for Theoretical and Experimental Physics,
B. Cheremushkinskaya 25, Moscow 117259, Russia}
\author{M. M\"uller-Preussker}
\affiliation{Institut f\"ur Physik, Humboldt-Universit\"at zu Berlin,
Newtonstr. 15, D-12489 Berlin, Germany}
\author{A. I. Veselov}
\affiliation{Institute for Theoretical and Experimental Physics,
B. Cheremushkinskaya 25, Moscow 117259, Russia}


\begin{abstract}
In equilibrium, at finite temperature below and above the deconfining phase 
transition, we have generated lattice $SU(2)$ gauge fields and have exposed 
them to smearing in order to investigate the emerging clusters of topological 
charge. Analysing in addition the monopole clusters according to the maximally 
Abelian gauge, we have been able to characterize part of the topological 
clusters to correspond either to non-static calorons or static dyons in the 
context of Kraan-van Baal caloron solutions with non-trivial holonomy. 
We show that the relative abundance of these calorons and dyons is 
changing with temperature and offer an interpretation as dissociation of 
calorons into dyons with increasing temperature. The profile of the Polyakov 
loop inside the topological clusters and the (model-dependent) accumulated 
topological cluster charges support this interpretation. 
Above the deconfining phase transition light dyons (according to Kraan-van 
Baal caloron solutions with almost trivial holonomy) become the most abundant 
topological objects. They are presumably responsible for the magnetic 
confinement in the deconfined phase. 
\end{abstract}


\maketitle


\section{Introduction}
\label{sec:introduction}
In our previous work~\cite{IMMPV05} we have started an investigation of 
topological objects that appear in the course of smearing~\cite{smearing} 
of equilibrium lattice fields. These were generated at finite temperature 
in the confining phase of SU(2) lattice gauge theory. The main idea was to 
analyze the monopole content of these objects, still far from being classical 
solutions, according to the maximally Abelian gauge (MAG) in order to 
classify them as related to non-static calorons or static dyons. 
Such a classification would be natural in the context of Kraan-van Baal 
(KvB) caloron solutions with non-trivial holonomy~\cite{KvB-1,KvB-2,LeeLu}. 
In the present paper, using some additional techniques, we take up this 
problem again, now for a couple of temperatures below and above $T_{\rm dec}$.

Before pointing out the new tools for this analysis, we should briefly recall 
the general view and the terminology used, that originate from the classical 
KvB solutions~\cite{KvB-1,KvB-2}. We will do this for the simplest case, the 
$SU(2)$ gauge theory. In this case, at finite temperature, the classical 
carriers of one unit of topological charge are selfdual or antiselfdual objects 
(periodic caloron solutions) which consist of {\it two} monopole ``constituents''.
In the limit of small inter-monopole distances they form a single, {\it non-static} 
caloron, whereas at large distances they dissociate into a pair of separate 
{\it static} lumps of action and topological density (static 
Bogomolnyi-Prasad-Sommerfield (BPS) monopoles \cite{BPS} or ``dyons'').  
In the latter case the action $8 \pi^2/g^2$ (or one unit of topological charge) 
is shared among the constituents according to the holonomy of the gauge field, 
in the ratio $~\overline{\omega}/\omega$, ~where 
$~\overline{\omega} = 1/2 - \omega$. The asymptotic value of the Polyakov loop 
\bea \nonumber
L_{\infty} &=& \lim_{|\vec{x}|\to \infty} L(\vec{x}) \\ \nonumber
           &=& \lim_{|\vec{x}|\to \infty} 
               \frac{1}{2} ~\mathrm{tr}~{\bf P}~ 
               \exp\left( i \int_0^{b=1/T} A_4(\vec{x}, t)~dt \right) \\ 
           &=& \cos(2\pi\omega), ~~0 \le \omega \le 1/2\,,
\label{eq:holonomy}
\eea 
determines the holonomy parameter $\omega$ and its complement $\overline{\omega}$.

Each physical phase can be thought as a medium creating a certain $L_{\infty}$ 
as boundary condition for the calorons. For example, maximally non-trivial 
asymptotic holonomy, 
$~L_{\infty} = 0~$ or $~\overline{\omega}=\omega=1/4$, forces the constituents 
to carry equal topological charge $~\pm 1/2$. With the holonomy becoming more and 
more trivial with increasing temperature, $~L_{\infty} \to \pm 1~$, the constituents
become imbalanced. On the other hand, the constituents are distinguished by the 
local behavior of the Polyakov loop $~L({\vec x}) \approx \pm 1$ close to their 
positions. This is the characteristic signature, irrespectively whether the action 
forms two separated lumps or a single lump, the latter one with unit topological 
charge. 

\begin{figure*}[!htb]
\centering
\includegraphics[width=.45\textwidth]{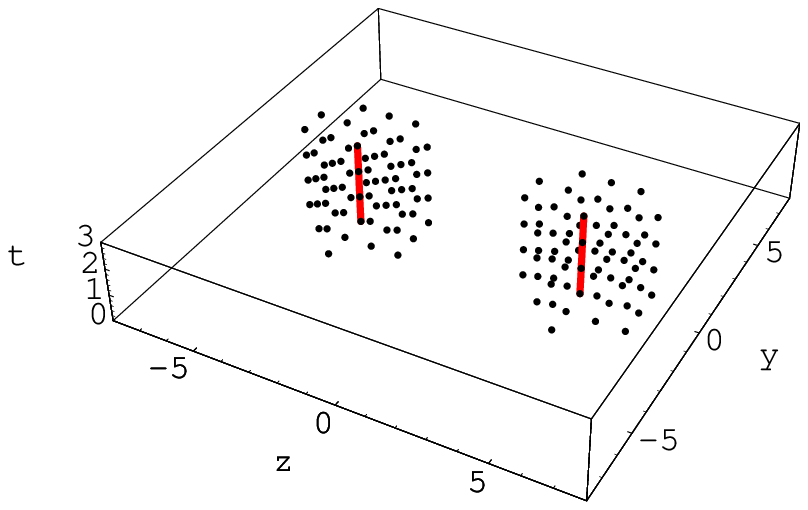}%
\hspace{0.5 cm}
\includegraphics[width=.45\textwidth]{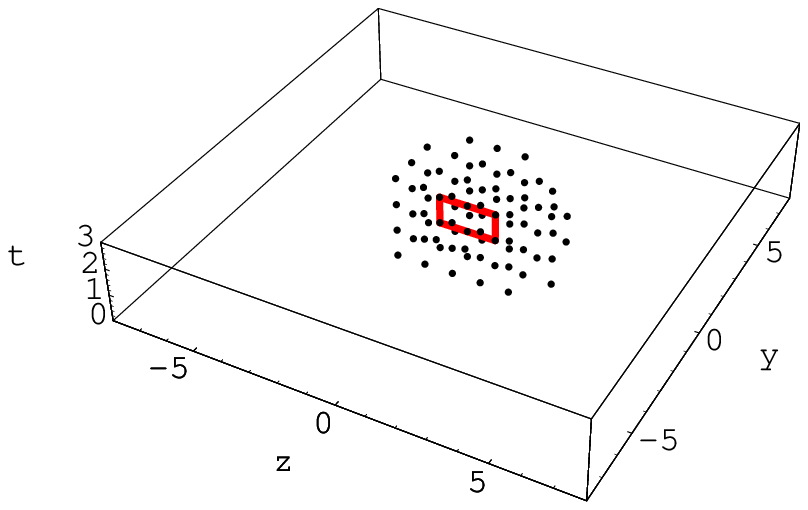}\\
\caption{(color online) Left: two static dyons accompanied by two static monopoles;
right: a single caloron accompanied by a monopole loop. 
The classical $SU(2)$ configurations have been generated on a $16^3 \times 4$ 
lattice and then Abelian-projected. 
In the $(z,y,t)$ lattice at $x=0$ we show the sites with action density 
$s>s_{\rm max}/5$ (left, for dyons) and $s>s_{\rm max}/40$ (right, for calorons)
together with the emerging monopole trajectories.}
\label{fig:example}
\end{figure*}

Viewed by Abelian monopoles appearing in the MAG, static dyons are represented 
by monopoles temporally wrapping around the lattice, whereas single non-static 
calorons are localized in time and have monopole loops of finite extent running 
close to the center. This feature of calorons in the MAG has been pointed out 
first by Brower et al.~\cite{Brower}. This was our guiding hypothesis already in 
Ref.~\cite{IMMPV05}. In Fig. \ref{fig:example} we show a classical caloron solution 
in the two limiting cases together with the corresponding MAG monopole content.
In Ref.~\cite{recombination} we have seen such static monopoles also in the 
Polyakov gauge, regularly accompanying static, {\it i.e.} dissociated caloron 
configurations while being rarely found in non-static calorons.

According to the scenario to be outlined in \Sec{sec:picture} we expect to 
observe below $T_{\rm dec}$ both dissociated and non-dissociated clusters of 
topological charge $\pm \frac{1}{2}$ and $\pm 1$, respectively, in some mixture. 
Above $T_{\rm dec}$ only clusters with charge $\pm 1$ should remain, with a 
density decreasing with increasing temperature. That means that we are expecting 
to see calorons with trivial holonomy actually dominating the topological charge 
above $T_{\rm dec}$.

The topological structure in the deconfined phase is partly also a question 
concerning the capability of the smearing method applied in this paper. Whereas 
the topological structure is known to be preserved under smearing in the confined 
phase, experience shows that under comparable smearing conditions the topology 
is rapidly wiped out in the deconfined phase. Therefore, in order to escape these 
bad prospects, we have to modify the smearing conditions in the deconfined phase.

We point out in \Sec{sec:picture} that there are fermionic methods to determine 
the topological density but, in order to detect a semiclassical structure, 
certain UV smearing techniques have to be applied as well.
Our analogous method applied for the same purpose is a combination of gluonic 
measurements of the most naive topological charge density combined with the 
monopole localization in the maximally Abelian gauge. Both are applied to APE 
smeared configurations. In a very recent paper it has been demonstrated, by 
comparison of smeared and cooled configurations with the original equilibrium 
configurations, that the spectrum of a chirally improved Dirac operator remains 
practically unaltered~\cite{GattringerIlgenfritzSolbrig} meaning that the infrared 
structure ``seen by the low-lying chiral fermion modes'' in equilibrium, is 
conserved under controlled cooling or moderate smearing.

Since for our purpose the correlation with Abelian monopoles is of crucial 
importance, the technical improvements (compared to Ref.~\cite{IMMPV05}) in 
the present paper are concerning the methods detecting the {\it monopole structure}.
This applies both to gauge fixing and to the analysis of separable monopole loops.  
We have now fixed the MAG by the more advanced technique of simulated 
annealing~\cite{simulated_annealing}. This method is known to lead, in comparison
with the overrelaxation method, to higher maxima of the gauge functional and to a 
minimal monopole content. Analogously to the preceding paper, the detection of 
monopole trajectories is used here to identify clusters of the topological charge 
density resembling the two extreme appearances of KvB calorons.  
Unlike the previous paper, we are now actually exploring the full monopole content 
of individual topological objects, not only looking for intersections of time-like 
monopole currents with the topological charge clusters~\cite{IMMPV05}. 
We are taking the precise {\it type of monopole clusters} into account.

Last but not least, we use now a larger lattice (with size $24^3\times 6$) 
and a certain set of different $\beta$-values. Three values, $\beta = 2.2,~2.3,~2.4$,
belong to the confined phase while the remaining two values of $\beta = 2.5,~2.6$, 
represent the deconfined phase. This will allow us to get information about the 
temperature dependence of the caloron/dyon composition that was left open in the
previous investiagtions.

In \Sec{sec:picture} we will outline the general physical picture behind the present 
investigation. In \Sec{sec:smearing} we will briefly explain the set of our runs
and in particular the conditions for the smearing procedure. 
In \Sec{sec:polyakov} we shall describe the local influence of monopoles altering 
the distribution of the Polyakov loop. This part is interesting by its own and is 
not related to the following cluster analysis. In \Sec{sec:cluster} we will describe
the results of the analysis of topological clusters. The classification with the 
help of monopoles will be of use here. We draw our conclusions in 
\Sec{sec:conclusion}.

\section{The physical picture}
\label{sec:picture}

The one-loop calculation of the caloron amplitude~\cite{Diakonov} has shown 
that below some temperature calorons may become instable with respect to the 
separation into their constituents. It would be attractive to identify this 
temperature with the deconfinement temperature because, in a dilute gas calculation,
it has been shown that at this temperature trivial holonomy $L_{\infty} \approx 1$ 
turns from a minimum of the free energy $P$ (as a function of $L_{\infty}$) into 
a maximum. The global view of the function $P(L_{\infty})$ is, however, beyond the 
capability of the {\it dilute caloron gas} approximation. In a schematic {\it dyon 
gas model}, Diakonov~\cite{Diakonov_review} has demonstrated how the stability of 
$L_{\infty}=0$ (being the minimum of $P$) could eventually be achieved. 

All this taken together leads to a simplified picture that describes the essence 
of the transition to confinement as the dissociation of tightly bound calorons 
(with trivial asymptotic holonomy) into dissociated ones with nontrivial holonomy 
and discernible dyonic constituentis. This cannot be the full truth, however. 
As we know from Refs.~\cite{Brower,recombination}, at low temperature calorons, 
even of nontrivial holonomy, are likely to appear as single lumps of action that 
nonetheless {\it are different from the well-known instanton solutions} because 
of the nontrivial holonomy boundary conditions they must satify in the confining 
medium. 

Therefore we expect to observe in a study at different temperatures the constituents
becoming manifest inside {\it dissociated calorons} more abundantly approaching 
the deconfining transition from below. Whereas the topological susceptibility 
is temperature independent throughout the confined phase, the relative abundance 
of static objects with an appreciable fractional topological charge (close to 
$\pm \frac{1}{2}$) should increase towards the phase transition temperature 
compared to the abundance of non-static calorons. 

We have announced in the Introduction that we have used four-dimensional 
smearing for this investigation. There is an alternative way to locate calorons 
and their constituents, respectively. It is based on the localization behavior 
of the zero modes and the near-zero modes of the Dirac operator \cite{GPGAPvB,CKvB}.
A jumping behavior of the zero mode in charge $Q=\pm1$ configurations can be 
observed under the influence of changing boundary conditions, deliberately imposed 
on the fermion field. This is very reminiscent to what is familiar from the case 
of an exact caloron background field. Interpolating with an appropriate phase factor
between periodic boundary conditions and antiperiodic ones, the zero mode may jump 
from one constituent to another. 

For various {\it classical} solutions extracted on the lattice by the cooling 
method this has been established for $SU(2)$~\cite{WeWithStanislav} as well as 
for $SU(3)$~\cite{WeWithPeschka}. For equilibrium lattice gauge fields selected 
to possess unit topological charge, in the $SU(3)$ and $SU(2)$ cases a similar 
``jumping'' of lowest-lying modes of a chirally improved lattice Dirac operator 
was reported by Gattringer and 
coauthors~\cite{GattringerSchaefer,GattringerPullirsch,GattringerSolbrig}.  
Of course, $Q=\pm1$ equilibrium configurations generically contain more clusters 
and spatial fluctuations of topological charge than the single zero-mode is able 
to detect. Such configurations differ from a pure (anti-)caloron even in the 
deconfined phase. It has been verified~\cite{Lattice03}, that the single zero mode 
is always attracted to positions that are maxima of the topological density inside 
extended topological clusters. This check has required a very well-tuned smearing 
and has used a topological charge operator defined in terms of the naive twisted 
plaquette. However, there were more clusters and cluster maxima going undetected
by the single zero mode. In our previous~\cite{IMMPV05} and in the present paper, 
the diagnostic role is shifted to the MAG monopoles. Note that these are not limited
in number or to a particular chirality as the zero modes are.

With a chirally perfect or improved Dirac operator at hand (that satisfies 
the Ginsparg-Wilson equation), there is a definition~\cite{overtopdefinition} 
of the topological charge density which is equivalent, in the continuum limit,
to the gluonic definition. It may supersede a gluonic definition at finite 
lattice spacing. This density also exposes a clustering 
pattern~\cite{Koma,Weinberg,Nicosia} strongly dependent on ultraviolet filtering, 
usually implemented by a cutoff (in eigenvalues) of the contributing eigenmodes. 
The non-filtered density, in contrast, contains fluctuations of all scales and 
cannot be related to a semiclassical model. An analysis of calorons and dyons using 
these tools along the ideas of the present paper is left for the future. 

In Ref.~\cite{InstantonQuarks} it has been attempted to construct single-caloron 
and multi-caloron solutions at lower temperature starting from high temperature 
ones. It was characteristic that all classical constituents were delocalized to 
a scale set exclusively by the periodic 4-volume, in contrast to the (jumping) 
zero modes seen also in zero temperature equilibrium configuration background 
configurations which are highly localized due to quantum fluctuations of the 
gauge field. This limits our expectations to find, in the limit of vanishing 
temperature, a semiclassical structure that would reveal the elusive ``instanton 
quarks'' in separation.

Using the new procedures under wider physical conditions at finite temperatures,
similar to Ref.~\cite{IMMPV05} we will observe both non-static calorons and static 
dyons. In the confined phase the relative population of isolated dyons will be seen
growing with {\it increasing} temperature. This will be interpreted as progressive 
dissociation (dipole splitting) of calorons into dyons in the confined phase. 
This is the main result of our paper directly related to the KvB caloron picture 
and confirming the simplified picture described above. 

Unexpectedly from the point of view of the scenario outlined above, in the 
deconfined phase nondissociated calorons are rare objects, even though charge 
$Q=\pm1$ configurations are still observed after a well-tuned amount of smearing. 
According to the different asymptotic holonomy (changing away from vanishing 
Polyakov loop toward $L_{\infty} \ne 0$) one would expect an extreme asymmetry 
among the constituent dyons if they appear at all. One type of dyons should be 
small in size and heavy in action with a peak of the Polyakov loop of opposite 
sign with respect to $L_{\infty}$, and another type of dyons large in size and 
light in action with a peak of the Polyakov loop of equal sign with the overall 
holonomy. The heavy dyons should be as rare as the nondissociated calorons are,
but they are not found paired with the light dyons. The latter will be found 
about an order of magnitude more frequently and contribute less to the topological 
charge. Being static and always connected with a static Abelian monopole, 
they could be held responsible (together with other monopoles) for the magnetic 
confinement in the deconfined phase. This is the second main observation to be 
reported.

\section{Details of the simulations and of smearing}
\label{sec:smearing}
We have generated 200 independent Monte Carlo configurations for each 
$\beta = 2.2,~2.3,~2.4,~2.5,~2.6$ on a $24^3\times 6$ lattice. We used the
heatbath method according to the Wilson plaquette action. These samples 
characterize finite temperature both in the confined and deconfined phases. 
At $N_{\tau}=6$, the critical $\beta$ corresponding to the deconfining phase 
transition is about $\beta_{dec} = 2.42$. 
For each $\beta$ the corresponding ratio $T/T_{\rm dec}$ is shown in Table I.
The ensembles further have been subjected to smoothing by four-dimensional 
APE smearing~\cite{smearing}. The fixed smearing parameter was $\alpha = 0.45$, 
whereas various numbers $N$ of iterations were applied at different $\beta$. 
From our previous observations on a $20^3\times 6$ lattice at $\beta = 2.3$ 
we concluded that dyons become visible above the noisy background after $N=50$ 
smearing steps.
\vspace{0.5cm}

\begin{figure}[!htb]
\begin{center}
\includegraphics[width=.3\textwidth,height=.3\textwidth]{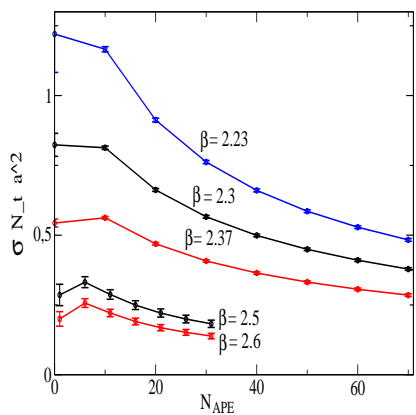}
\caption{(color online) The effect of smearing on the ``string tension'' $\sigma$.
In confinement ($\beta = 2.23,~2.3,~2.37$) $\sigma$ was obtained from the 
correlator of Polyakov loops. In deconfinement ($\beta = 2.5,~2.6$) $\sigma$ was 
obtained from the behavior of space-like Wilson loops. The lattice size was 
$20^3\times 6$.}
\label{fig:sigma}
\end{center}
\end{figure}

For the present investigation at $\beta = 2.3$ and the two other $\beta = 2.2,~2.4$ 
in the confined phase we adopt this number of smearing steps, justified by the 
observation that the string tension~\footnote{For the purpose of this estimate the 
string tension was defined by the Polyakov loop correlator.} at the neighbouring 
values $\beta = 2.23,~2.3,~2.37$ is reduced after $50$ smearing steps to 
approximately the same $60 \%$ of the original value (see \Fig{fig:sigma}).
In the deconfined phase, where the {\it spatial} string tension 
was used for this comparison, the same percentual reduction of this string tension 
has been observed at $N=25$ and $N=20$ smearing steps for $\beta = 2.5,~2.6$, 
respectively.

\section{Local Polyakov loop, monopoles and asymptotic holonomy}
\label{sec:polyakov}
\begin{figure*}[!htb]
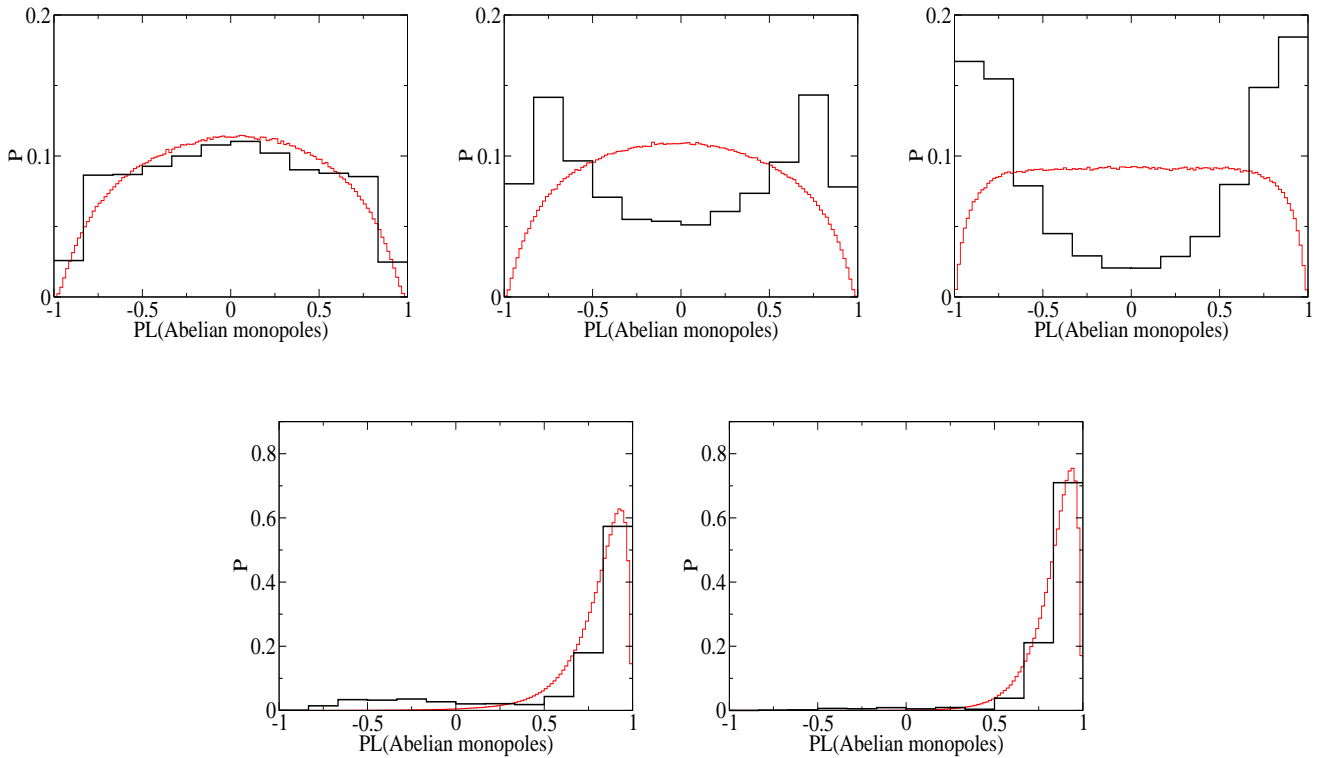

\centering
\includegraphics[width=.3\textwidth,height=.25\textwidth]{fig3a.eps}%
\hspace{0.5 cm}
\includegraphics[width=.3\textwidth,height=.25\textwidth]{fig3b.eps}%
\hspace{0.5 cm}
\includegraphics[width=.3\textwidth,height=.25\textwidth]{fig3c.eps}\\
\vspace{1 cm} 
\includegraphics[width=.3\textwidth,height=.25\textwidth]{fig3d.eps}%
\hspace{0.5 cm}
\includegraphics[width=.3\textwidth,height=.25\textwidth]{fig3e.eps}\\
\caption{(color online) The distribution of the Polyakov 
loop over sites where time-like 
Abelian monopole currents are detected (thick line). Monopoles are obtained 
by Abelian projection in MAG. For comparison, the distribution of Polyakov 
loops over all sites is shown (thin line).
The upper row of figures shows the results for $\beta = 2.2,~2.3,~2.4$ 
(from left to right, confinement). The lower row of figures shows the results 
for $\beta = 2.5,~2.6$ (from left to right, deconfinement).}
\label{fig:plmon}
\end{figure*}

Our first observation, as in Ref.~\cite{IMMPV05}, is the correlation between 
the values of the Polyakov loop and the presence of Abelian monopoles. 
For this purpose we have averaged the Polyakov loop over all eight corners of
a three-dimensional cube (dual to a time-like link of the dual lattice) where 
an Abelian magnetic charge has been detected. This construction is implied  
whenever the original and the dual lattice are put into relation. The distribution 
of the values of this conditionally averaged Polyakov loop is shown in the different
panels of Fig. \ref{fig:plmon} as histograms (with broad 12 bins, drawn in thick 
lines). For comparison, the distributions of the Polyakov loop over {\it all} 
lattice sites are shown (as fine-binned histograms, drawn in thin lines). 
One sees that, with $\beta$ approaching the phase transition from the confinement 
side, the distribution over all lattice sites becomes flatter, whereas the 
distribution over the position of monopoles becomes increasingly shifted 
toward $\pm 1$. 

Within our set of runs above but not so close to the deconfinement temperature
global flips of the Polyakov loop did not happen. Due to this effective $Z(2)$ 
breaking the two plots for the deconfined phase show an unsymmetric distribution 
for the Polyakov loop over all lattice sites. At these temperatures the same 
asymmetry is seen in the distribution of the Polyakov loops for the monopole 
positions.

\begin{figure*}[!htb]
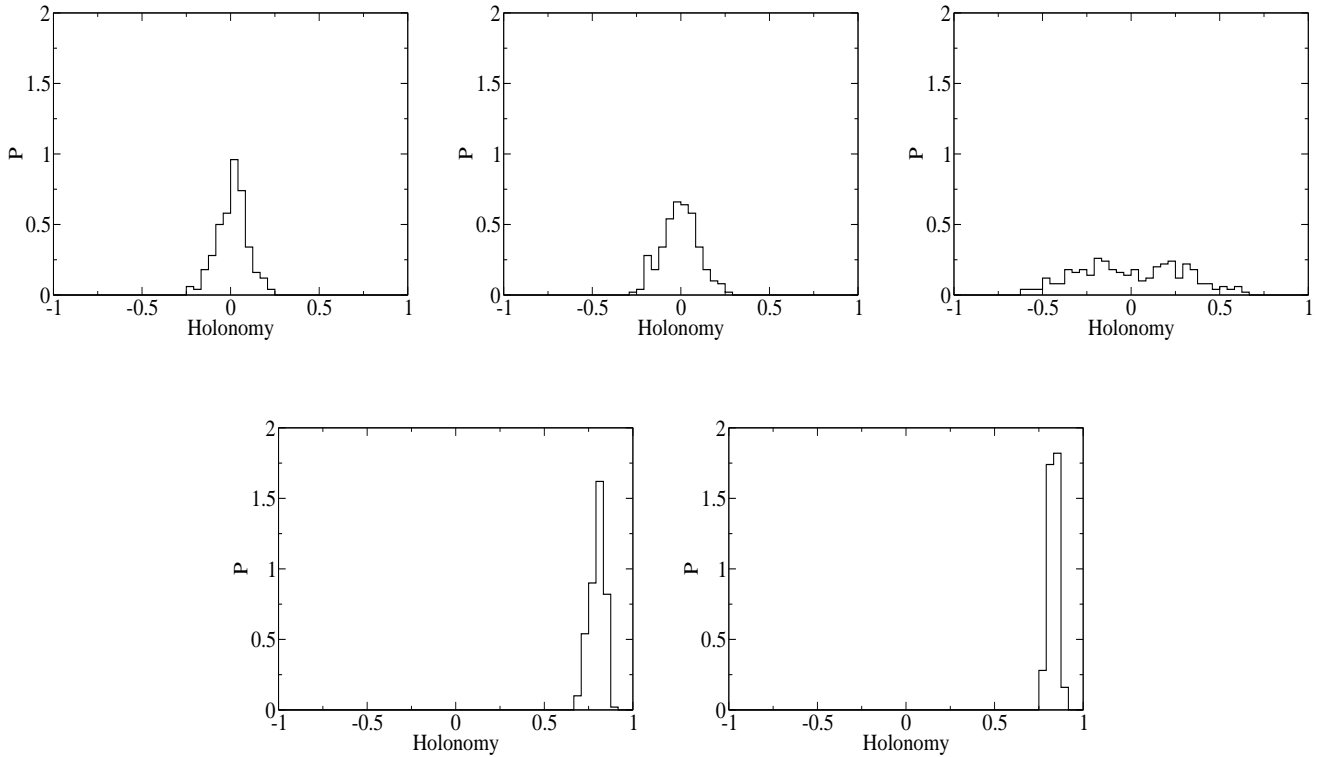

\centering
\includegraphics[width=.3\textwidth,height=.25\textwidth]{fig4a.eps}%
\hspace{0.5 cm}
\includegraphics[width=.3\textwidth,height=.25\textwidth]{fig4b.eps}%
\hspace{0.5 cm}
\includegraphics[width=.3\textwidth,height=.25\textwidth]{fig4c.eps}\\
\vspace{1 cm}
\includegraphics[width=.3\textwidth,height=.25\textwidth]{fig4d.eps}%
\hspace{0.5 cm}
\includegraphics[width=.3\textwidth,height=.25\textwidth]{fig4e.eps}\\
\caption{The distribution of the ``asymptotic'' 
holonomy $H$ as defined in the text. The upper row of figures shows 
the results for $\beta = 2.2,~2.3,~2.4$ (from left to right, confinement).
The lower row of figures shows the results for $\beta = 2.5,~2.6$ (from left 
to right, deconfinement).}
\label{fig:holonomy}
\end{figure*}

In order to facilitate a qualitative comparison of the topological clusters with
KvB caloron solutions with nontrivial holonomy $L_{\infty}$, we have defined 
an empirical {\it asymptotic holonomy} $H$ for each lattice configuration. We 
have considered the sites on the lattice where the absolute value of topological 
charge density is {\it less than the averaged absolute value} of this density. 
We take the average of the Polyakov loop over this set of asymptotic sites and 
call the resulting observable the asymptotic holonomy $H$. The distribution of 
$H$ for the five $\beta$ values is shown in Fig. \ref{fig:holonomy}.

It can be seen from the different panels of this figure that the asymptotic 
holonomy $H$ is concentrated near zero deep in the confined phase ($\beta =2.2$
and 2.3) . The distribution is widened at $\beta=2.4$ , already close to the 
deconfining transition, and it becomes again concentrated around a maximum moving 
slowly between $+0.5$ and $+1$ for our temperature range in the deconfined phase.

Note the sharp contrast characteristic for the confinement phase of the $H$ 
distribution to the Polyakov loop distribution at the positions of time-like 
monopoles (as defined in MAG) in Fig. \ref{fig:plmon}. In the deconfined phase 
the local Polyakov loop at the monopole positions is distributed around the 
asymptotic (global) holonomy $H$.

\section{Cluster analysis of smeared configurations}
\label{sec:cluster}
On one hand, for each of the four-dimensionally smeared configurations
we have looked for clusters of topological charge. The topological density 
is assigned to the lattice sites according to the unimproved (naive) twisted
plaquette definition. In order to form clusters for the subsequent analysis 
we have selected the sites where the absolute value of the topological charge 
density exceeds some threshold value $q_c$. The link-connected sites with 
$q(x) > q_c$ or $q(x) < - q_c$ form what we call positive (negative) clusters 
of topological charge. Obviously, number and size of the clusters depend on the 
threshold $q_c$. This value has been varied between the ensemble average of 
$|q(x)|$, taken as lowest threshold, and a value taken $10$ times larger taken
as the highest threshold with the aim to get, for each smeared configuration, 
a maximal number of mutually disconnected clusters of topological charge.
Therefore, this procedure finds an upper bound for the number of clusters
per configuration (corresponding to the degree of smearing that we have chosen).
This number can be inferred from Table I and amounts to 20 to 55 clusters
per configuration, maximal at the lowest and highest temperature, minimal 
at $T \lesssim T_c$. Obviously, the number of clusters is not a well-defined
quantity, and many of the shallow clusters should be ignored as mere extended 
background fluctuations. 

\begin{table}[ht]
\caption{The numbers of topological clusters for $200$ smeared configurations
at different $\beta$ identified as static dyons (D), non-static calorons (CAL) 
and without clear identification (OTHER).}
\begin{center}
\begin{tabular}{lcccc}
\hline
~~~ $\beta$~~~ & $T/T_{\rm dec}~~~$ &  $D~~~$ & $ CAL~~~$ & $OTHER~~~$ \\
\hline
~~~ $2.2~~~$ & $0.54~~~$ & $ 39~~~ $  & $ 624~~~ $ & $ 10323~~~ $   \\
\hline
~~~ $2.3~~~$ & $0.68~~~$ & $ 128~~~ $ & $ 755~~~ $ & $ 6512~~~ $   \\
\hline
~~~ $2.4~~~$ & $0.94~~~$ & $ 626~~~ $ & $ 511~~~ $ & $ 2983~~~ $   \\
\hline
~~~ $2.5~~~$ & $1.32~~~$ & $ 243~~~ $ & $  84~~~ $ & $ 6989~~~ $   \\
\hline
~~~ $2.6~~~$ & $1.87~~~$ & $ 118~~~ $ & $  20~~~ $ & $ 10204~~~ $   \\
\hline
\end{tabular}
\end{center}
\end{table}

On the other hand, we have identified the complete monopole structure for each 
smeared configuration. This means that all monopole clusters have been found. 
A monopole cluster is a connected set of occupied links of the dual lattice, 
{\it i.e.} all links carrying non-vanishing magnetic charge current. The density 
of monopole currents ({\it i.e.} the percentage of occupied links of the dual 
lattice) in the smeared configurations was about two orders of magnitude smaller 
than in the corresponding equilibrium ones. Therefore, in the smeared ensemble 
monopole clusters were mainly non-selfintersecting loops, in the confinement 
phase yet percolating loops.
From the set of all monopole clusters in a configuration, we have selected those 
{\it isolated monopole loops} 
that are either closed by periodicity in the time direction (then mainly formed 
by minimal length $L=6$ static time-like monopole currents) or closed within the 
four-dimensional volume (then mainly with the minimal loop sizes $L=4$, $6$, $8$). 
If there was a complete covering of exactly one monopole loop (of either kind) 
with some topological cluster, we have classified this cluster either as a static 
dyon or as a nondissociated caloron, which are the two opposite appearances of 
KvB calorons. Table I shows the corresponding numbers of topological clusters 
classified in one or the other way. Note that our classification, due to the very 
restrictive cut, has been passed only by a small fraction of topological clusters.
The number of topological clusters in general is best defined closely below the 
phase transition ($\beta=2.4$) where it reaches a minimum of $\approx 15$ per 
configurations. 
Approximately one third of them becomes successfully classified as dyons or 
calorons, the rest has no obvious interpretation in terms of the monopole content. 
The minimal monopole clusters mentioned above could be intersecting with the large, 
percolating monopole loops, hence having escaped our attention.  

With the 4-dimensional volume of $17.84 \mathrm{~fm}^4$ (at $\beta=2.4$ the lattice 
spacing is obtained from $\sigma a^2=0.071(1)$~\cite{LuciniTeper}) the density of 
identified calorons could be estimated as $0.143 \mathrm{~fm}^{-4}$, the density
of dyons $0.175 \mathrm{~fm}^{-4}$. The relative abundance extrapolated to the full 
number of clusters at this $\beta$ would for the calorons result in  
$0.374 \mathrm{~fm}^{-4}$, still in the right ballpark given the constant topological
susceptibility $\chi \approx 1 \mathrm{~fm}^{-4}$ below $T_{\rm dec}$. 
Expressed in physical units, one can summarize the outcome as follows:
the density of monopoles (dyons) continuously increases with the temperature.
Near the phase transition the densities of calorons and monopole-like objects
become comparable. We will discuss below to what extent the monopole-like objects 
are really dyons and may be described as caloron constituents. The density of 
objects classified as calorons drops down above the phase transition. This does
not come not unexpected in view of the topological susceptibility decreasing with 
increasing temperature above $T_{\rm dec}$. It remains to be seen 
whether, in the deconfined phase, the configurations with topological charge 
$Q=\pm1$ really contain such a caloron. This would have to be expected from the 
scenario that calorons are mainly undissociated in the deconfined phase.

\begin{table}[ht]
\caption{The densities (in fm$^{-4}$) of dyon and caloron clusters, as identified 
(in the second line extrapolated to the full number of topological charge clusters).}
\begin{center}
\begin{tabular}{lccc}
\hline
~~~ $\beta$~~~ & $T/T_{\rm dec}~~~$     & $D~~~$              & $ CAL~~~$  \\
\hline
~~~ $2.2~~~$   & $0.54~~~$ &            $ 1.14\cdot 10^{-3}$  & $ 0.0182$    \\
~~~ $2.2~~~$   & $0.54~~~$ &            $ 0.018            $  & $ 0.283$     \\
\hline
~~~ $2.3~~~$   & $0.68~~~$ &            $ 9.78\cdot 10^{-3}$  & $ 0.0577$    \\
~~~ $2.3~~~$   & $0.68~~~$ &            $ 0.062            $  & $ 0.425$      \\
\hline
~~~ $2.4~~~$   & $0.94~~~$ &            $ 0.175            $  & $ 0.1429$      \\
~~~ $2.4~~~$   & $0.94~~~$ &            $ 0.458            $  & $ 0.374$       \\
\hline
~~~ $2.5~~~$   & $1.32~~~$ &            $ 0.260            $  & $ 0.0900$      \\
~~~ $2.5~~~$   & $1.32~~~$ &            $ 5.55             $  & $ 1.92$      \\
\hline
~~~ $2.6~~~$   & $1.87~~~$ &            $ 0.514            $  & $ 0.0872$      \\
~~~ $2.6~~~$   & $1.87~~~$ &            $ 38               $  & $ 6.45$      \\
\hline
\end{tabular}
\end{center}
\end{table}

In order to corroborate (or relativize, for the deconfined phase) this 
interpretation we have tried to
characterize the monopole-tagged clusters by some cluster variables.
We have averaged the Polyakov loop inside the selected clusters again over all 
those sites where time-like Abelian monopole currents of either sign are observed 
as part of the closed monopole loop that served the identification. 
We call this quantity the Polyakov loop averaged over the ``monopole skeleton'' 
of the given cluster,
$\langle PL(\mathrm{Abelian~monopoles})\rangle_{cluster}$. 
From KvB calorons we expect this average for ``static monopole clusters'' 
to be close to $~\pm 1~$, whereas for ``monopole--antimonopole pair clusters''
it should be close to $~0~$, the latter because of the internal dipole structure 
in terms of the Polyakov loop that reflects the positions of the constituents 
which otherwise would be hidden inside the single undissociated caloron cluster.  
We use this as the first cluster variable that should quantitatively characterize 
the type of cluster.  

As a second cluster variable we want to consider the topological charge 
assigned to the cluster. This total charge is difficult to assess because 
the topological profile extends below the threshold $q_c$. Applying this cutoff
was, however, indispensable to localize the cluster. We have designed a 
{\it model-dependent} estimator for the total topological charge of an identified 
topological cluster.
\begin{figure*}[!htb]
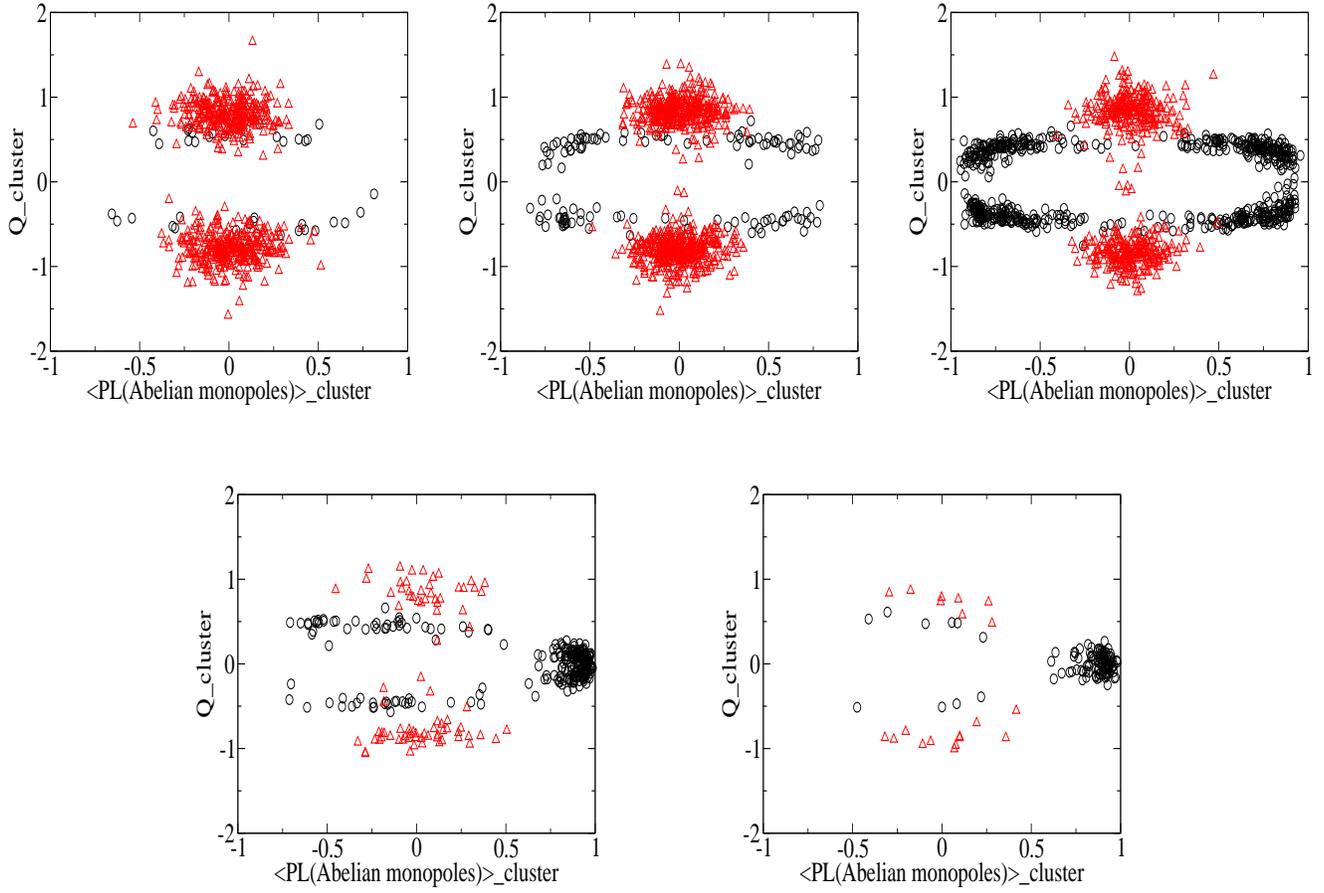

\centering
\includegraphics[width=.3\textwidth,height=.3\textwidth]%
{fig5a.eps}%
\hspace{0.5 cm}
\includegraphics[width=.3\textwidth,height=.3\textwidth]%
{fig5b.eps}%
\hspace{0.5 cm}
\includegraphics[width=.3\textwidth,height=.3\textwidth]%
{fig5c.eps}
\\
\vspace{1 cm}
\includegraphics[width=.3\textwidth,height=.3\textwidth]%
{fig5d.eps}%
\hspace{1.5 cm}
\includegraphics[width=.3\textwidth,height=.3\textwidth]%
{fig5e.eps}
\\
\caption{(color online) Scatter plots in the
($Q_{cluster}$,$~\langle PL(\mathrm{Abelian~monopoles})\rangle_{cluster}$) 
plane for $\beta=2.2,~2.3,~2.4$ (upper row from left to right, confinement) 
and for $\beta=2.5,~2.6$ (lower row from left to right, deconfinement). 
The circles represent dyon clusters, the triangles undissociated calorons.}
\label{fig:plqcorr}
\end{figure*}
This estimator works differently for the two types of clusters.
For a static dyon we know from the analytic KvB caloron solution that its 
size depends on the holonomy parameter $\omega$ according to \Eq{eq:holonomy} 
as
\begin{equation}
\bar{r} = \frac{b}{4\pi\omega}~.
\label{eq:dyonsize}
\end{equation}
$b$ is the inverse temperature, i.e. the period in time direction.
In the confinement phase assuming the holonomy parameter to be 
$\omega = 0.25$ the size of the dyons changes inversely to temperature.
Actually, the sizes of the dyon clusters are distributed within some width.
According to the caloron solution in the limit of separated dyons,
the modulus of the topological charge density $q_{max}$ in the center 
$x_{max}$ of the dyon cluster should be connected with the cluster size 
in the following way:
\begin{equation}
q_{max} = \frac{1}{24\pi^2\bar{r}^4}~.
\label{eq:dyondens}
\end{equation}
Ascribing by this relation a size to each cluster classified as static dyon
we could obtain a cluster size distribution from the observed maxima of the 
topological charge density inside the clusters.
Then, considering such a cluster, we sum the actual topological charge over 
all sites $x$ that belong to the cluster including the tail below $q_c$. 
That means that we sum the topological charge density $q(x)$ within a tube 
with a {\it 3-dimensional} (spatial) radius $R$ (related to $~\bar{r}$).  
This distance $R$ should not be too large in order to avoid double counting 
of topological charge (by assigning sites to more than one cluster) 
and  not too small (in order not to underestimate the topological charge in
the tail of the cluster under consideration). We use $R=3~\bar{r}$ which, 
for the case of an isolated dyonic cluster (with an ideal, approximately 
exponential topological charge density profile), would estimate the total 
charge within $7\%$ accuracy. This estimator corrects for the tail and 
serves to assign a topological charge to all clusters once they have been 
classified as static dyon clusters.

An undissociated KvB caloron has a topological charge profile like that of
an isolated ordinary instanton solution. In this case the maximum of the 
modulus of the topological charge density is related to the instanton size 
$~\rho~$ as follows
\begin{equation}
q_{max} = \frac{6}{\pi^2\rho^4}~.
\label{eq:instdens}
\end{equation}
Assuming that the clusters classified as undissociated calorons have such a 
charge profile we can obtain the instanton size $~\rho~$ from the measured 
$~q_{max}~$ of the cluster. Then we sum the actual topological charge density 
over all sites $x$ inside a {\it 4-dimensional} ball with a radius $1.5\rho$ 
centered at $x_{max}$. Finally, the result needs to be multiplied by a 
correction factor $~1.29~$ inferred from the exact instanton solution. 
In this way, we are in the position to define an estimated topological charge 
for any cluster once it has been identified as undissociated caloron.

\begin{figure*}[!htb]
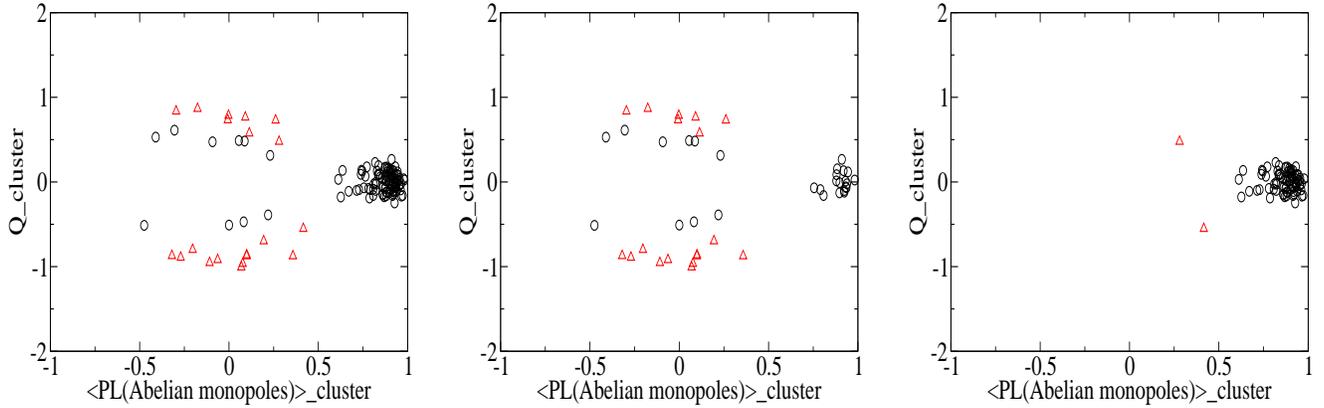

\begin{center}
\vspace*{0.5cm}
\includegraphics[width=.3\textwidth,height=.3\textwidth]%
{fig6a.eps}%
\hspace{0.5 cm}
\includegraphics[width=.3\textwidth,height=.3\textwidth]%
{fig6b.eps}%
\hspace{0.5 cm}
\includegraphics[width=.3\textwidth,height=.3\textwidth]%
{fig6c.eps}\\
\caption{(color online) Scatter plots in the
($Q_{cluster}$,$~\langle PL(\mathrm{Abelian~monopoles})\rangle_{cluster}$) 
plane for $\beta=~2.6$. The left figure is the sum of the other two.
The figure in the center shows configurations (45 from 200 configurations) 
with topological charge $|Q| = 1$ ($0.5\le |Q| \le 1.5$),
the right figure the complementary 155 configurations with zero topological 
charge $Q = 0$ ($|Q| \le 0.5$). The meaning of the symbols is the same as in 
Fig. \ref{fig:plqcorr}.  }
\label{fig:plqcorrq0q1}
\end{center}
\end{figure*}

We show in Fig. \ref{fig:plqcorr} scatter plots showing clusters of our
topological clusters in the plane spanned by the estimated topological 
charge of each cluster, $~Q_{cluster}~$, and the Polyakov loop averaged 
over the monopole skeleton, denoted as
$\langle PL(\mathrm{Abelian~monopoles})\rangle_{cluster}$.
The points in these scatter plots represent cluster that have been identified 
in one or the other way. As it can be seen from this figure, in the confined 
phase with increasing $\beta$ (here at $\beta = 2.3,~2.4$) the two sorts of 
topological clusters are forming clusters on the scatter plot either closer 
to the points 
$~\langle PL(\mathrm{Abelian~monopoles})\rangle_{cluster} =\pm c, 
~Q_{cluster} = \pm 1/2~$ (dissociated) or 
$~\langle PL(\mathrm{Abelian~monopoles})\rangle_{cluster} = 0,
~Q_{cluster} = \pm 1~$ (undissociated),
with $c \to 1$ with increasing temperature. In the confined phase at lower 
temperature, $\beta = 2.2$, clusters which contain closed monopole loops 
(``undissociated calorons'') are the most abundant objects (among the identified 
ones) and are forming a cluster, very similarly to $\beta = 2.3,~2.4$ .
On the other hand, dissociated dyons are very rare and have not yet produced a 
pronounced pattern of non-vanishing Polyakov loop (when the Polyakov loop is
averaged over the monopole skeleton of the cluster).

In the deconfined phase at $\beta = 2.5,~2.6$ undissociated calorons are
forming clusters in the same way as they do in the confined phase. However,
they become increasingly rare objects with increasing temperature. Single 
(dissociated) dyons are presented now by asymmetric objects: 
rare heavy dyons with an averaged Polyakov line of opposite sign compared to 
the holonomy $H$ and frequent light dyons with averaged Polyakov line
approximately equal to the value of the asymptotic holonomy. 
Undissociated calorons and heavy dyons are appearing almost exclusively in
configurations with total topological charge equal to $Q=\pm 1$ while  
light dyons appear mainly in configurations with total topological charge 
equal to $Q=0$ and to a lesser extent in $Q=\pm1$ configurations 
(see Fig. \ref{fig:plqcorrq0q1}). The charge values $Q=0, \pm 1$ exhaust 
all cases among the smeared configurations on the given lattice size, that
are found for the two $\beta$-values in the deconfined phase.
This suggests to view the topological background in the deconfined phase
mainly as a gas of ``spurious'' (both with small action and topological charge) 
dyon--antidyon pairs. These pairs cannot be understood as forming together a 
caloron. If they were related to calorons at all, more likely they have emerged 
from a caloron-anticaloron pair by annihilation of their heavy partners.
For these dyons it has to be checked whether they are formed by locally 
selfdual or anti-selfdual field. Modeling the size of these objects starting
from the topological density in the center is more questionable than in the
other cases.
Considered only as static Abelian magnetic monopoles, they do not have to follow
the dyonic interpretation. However they could be held responsible for the 
magnetic confinement. From this point of view it is no surprise that their
density increases with temperature above $T_{\rm dec}$. 

Under further cooling, pairs of dyons and antidyons in this gas can eventually
annihilate. In some cases (when the annihilation proceeds across the periodic 
boundary) this process ends up with Dirac sheets ({\it i.e.} constant magnetic 
fluxes)~\cite{Dirac_sheets_we}. These Dirac sheets are either stable or 
unstable depending on the value of holonomy~\cite{Stability_PvB}.

\begin{figure*}[!htb]
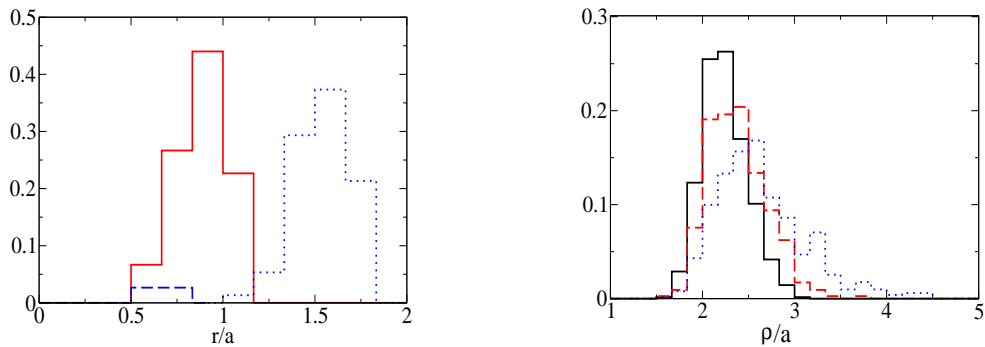

\begin{center}
\includegraphics[width=.3\textwidth,height=.25\textwidth]{fig7a.eps}
\hspace*{2cm}
\includegraphics[width=.3\textwidth,height=.25\textwidth]{fig7b.eps}
\caption{(color online) Left: distributions of the 
radii $\bar{r}$ of dyonic clusters  
for $\beta = 2.3$ (solid lines, confinement phase) and $\beta = 2.6$ 
(dashed/dotted lines, deconfinement phase). The dashed line corresponds 
to heavy dyon clusters, whereas the dotted line refers to light dyons.
Right: distributions of the scale-sizes $\rho$ of isolated caloron clusters 
for $\beta=2.2$ (solid line), $\beta=2.3$ (dashed line), and $\beta=2.4$
(dotted line). Both the values for $\bar{r}$ and $\rho$ are presented in 
units of the lattice spacing $a(\beta)$.}
\label{fig:radrho}
\end{center}
\end{figure*}
With some reservations concerning the dyonic interpretation of the monopole-like
clusters in the deconfined phase, we have made a comparison of the ``dyon'' radii 
found in the confined and deconfined phases. In the left half of \Fig{fig:radrho} 
we show the size distributions derived from the $q_{max}$-values of clusters 
classified as single dyons. For the confinement phase, this yields a distribution
with a single maximum as expected from the caloron model taken seriously.
In the deconfined phase, the discrimination between the abundant light 
dyon clusters with the same sign of the Polyakov loop at the monopole 
position and the rare heavy dyon clusters with opposite sign, always with 
respect to $H$, is reflected in the figure and highlighted by dotted and dashed 
lines, respectively. For comparison, we show in the right part of \Fig{fig:radrho} 
the $\rho$ distributions of isolated single caloron clusters found in the 
confinement phase. These distributions undoubtedly reflect the increasing size
of the objects yet to be classified as single calorons. This tendency fits
together with the increasing amount of dyons since, with increasing size,
calorons eventually turn into discernible dyons.

According to the relation (\ref{eq:dyonsize}),   
an estimate of $\omega = 0.25$ for maximally nontrivial holonomy 
at $\beta = ~2.3$, and $\omega = 0.1$ corresponding to 
$\langle H \rangle = 0.8$ at $\beta= ~2.6$ could be taken for rough orientation. 
Since the inverse temperature is $b=6a$ (with $N_{\tau}=6$) in all our cases, 
we would estimate $\bar{r} \approx 2 \times a(\beta)$ 
for all dyonic constituents in the confined phase,
whereas at $\beta= ~2.6$ in the deconfined phase the light dyons 
should have a typical radius $\bar{r} = ~4.8\times a(2.6)$ 
and the heavy dyons (with $\omega$ replaced by 
$\bar{\omega} = 0.5 - \omega = 0.4$) a radius of 
$\bar{r} = ~1.2\times a(2.6)$. 
Notice that the lattice spacings for the two $\beta$-values are widely 
different (they differ by a factor 2.75). 
Taking this into account
the dyons at $\beta = ~2.3$ (equally weighted in the confinement phase) 
would have a size of 
$\bar{r} \approx ~0.6/\sqrt{\sigma}$, that happens to be almost equal 
to the size of the light dyons at $\beta = ~2.6$ 
(in the deconfined phase) which is $\bar{r} \approx ~0.64/\sqrt{\sigma}$.
The heavier dyons at $\beta = ~2.6$ in the deconfined phase 
(which are actually rather rare) would be 4-times smaller in size, 
namely $\bar{r} \approx 0.16/\sqrt{\sigma}$. 
For the lattice spacings $a(\beta)$ 
assumed here in order to express all sizes in terms of the $SU(2)$ 
string tension at $T=0$ we refer to Ref.~\cite{LuciniTeper}.

Concluding we can say that the distributions drawn in \Fig{fig:radrho} 
do not agree quantitatively with the above estimates dictated by  
holonomy values $H \equiv L_{\infty}$, but show at least qualitatively 
the expected pattern of size splitting. 
On one hand, the averaged holonomy $H$ might not be a relevant parameter 
for the asymptotics of a single topological cluster.
On the other hand, we
would not take these caloron-based estimates too serious for the
static monopoles seen in the deconfined phase, because they 
are not observed in heavy-light pairs and are badly modelled by the
formula for calorons and selfdual BPS dyons.

\section{Conclusion}
\label{sec:conclusion}
We have analyzed the topological clusters of four-dimensionally smeared 
configurations at finite temperatures, across the deconfining phase transition,
by means of studying the monopole-worldline cluster content of these clusters.
This tool has allowed to identify part of the topological clusters either
as static dyons or non-static calorons. We have to stress, however, that
the number of separable topological clusters was maximized in the course of 
analysis such that a huge part of the clusters (more than 90 \% at the lowest 
temperature and in the deconfined phase, approximately 70 \% closely below the 
transition temperature) remained unidentified in this sense. This doesn't mean
that they couldn't be classified using other means. The simple monopole clusters
that were thought to identify clusters as dyons or calorons could have been 
intersecting with the percolating part that exists in the confinement phase.
Another part of the topological clusters, mainly in the deconfined phase,
should simply have been ignored because they are shallow background fluctuations. 
These have nothing to do whatsoever with the well-localized, (anti)selfdual 
carriers of topological charge which had to be analyzed.

On the basis of this separation we attempted a model-dependent reconstruction 
of the total topological charge for each individual cluster. The correlation 
between the Polyakov loop (averaged over the monopole skeleton) and the total 
topological cluster charge has revealed the following pattern. 
At temperatures not too much below the deconfining phase transition, with the 
asymptotic holonomy being still maximally nontrivial, $H=0$, we see both 
nondissociated calorons and separate dyons, understood as the result of 
part of the calorons being already dissociated into dyons. The size of the
nondissociated ones slightly increases with temperature.
Dyons with different local values of the Polyakov loop are symmetric in mass 
and action in the confinement phase. The ratio of the number of dyons to the 
number of (yet undissociated) calorons is increasing with temperature, such 
that we may speak about a process of dissociation with increasing temperature. 
At the transition temperature the densities of both objects has become comparable.

Above the deconfining phase transition holonomy starts to become trivial, 
$H \ne 0$. Contrary to our expectations topological excitations are {\it not 
dominated} by (undissociated or dissociated) calorons. 
Only half of the $Q=\pm1$ configurations can be interpreted as undissociated
calorons. Judged solely from the average Polyakov loop at the static clusters,
dyons are still prevailing above $T_{dec}$. The most abundant objects are 
light dyons and antidyons with static Abelian monopoles inside. 
That dyons are no more symmetric in mass and action would have been expected.
But they do not appear anymore as pairs of light and heavy constituents 
in the same configuration. Therefore it is difficult to understand these 
configurations as dissociated calorons.

The typical $Q=0$ configurations contains monopole--antimonopole pairs, each 
with only insignificant topological charge. 
Since the estimate of the charge rests on formulae for BPS dyons, some
reservations about the topological charge are due.
Considered simply as static magnetic monopoles (ignoring the topological 
charge cluster) together with other magnetic monopoles they 
could be held responsible for the spatial string tension 
(magnetic confinement) in the deconfined phase. From this point of view
it is plausible that the density of monopoles does not follow the decline
of the topological susceptibility above $T_{\rm dec}$.

The logical next step will be to repeat the present study with a suitable 
fermionic topological charge density applied to the (unsmeared) equilibrium 
gauge field configurations with the option of eliminating UV fluctuations 
by mode truncation.  

\vspace*{-0.5cm}
\section*{Acknowledgements}
This work was partly supported by RFBR grant 04-02-16079,  
DFG grant 436 RUS 113/739/0-2 and RFBR grant 06-02-16309.  
E.-M. I. is supported by DFG (FOR 465 / Mu932/2).
Two of us (B.V.M. and A.I.V.) gratefully appreciate the support of 
Humboldt-University Berlin where this work was carried out to a large extent. 
E.-M. I. thanks the Institute for Nuclear Theory at the University of
Washington for its hospitality and the Department of Energy for partial
support during the completion of this work. 
He wishes to thank the organizers of the Program INT-06-1, in particular
J. Negele and E. Shuryak, for the opportunity to discuss this work and 
related aspects of dyons in finite temperature QCD at Seattle.



\end{document}